\documentstyle[12pt]{article}

\title{Analytical Solution of the Off-Equilibrium Dynamics of a Long
Range Spin-Glass Model}
\vspace{3.5in}
\author{L. F. Cugliandolo and J. Kurchan \\
Dipartimento di Fisica, Universit\`a di Roma,
  {\em La Sapienza},\\
I-00185 Roma, Italy\\
and\\
INFN Sezione di Roma I, Roma, Italy}
\vspace{7in}

\begin{document}

\newcommand{\beq}{\begin{equation}}
\newcommand{\eeq}{\end{equation}}
\newcommand{\beqn}{\begin{eqnarray}}
\newcommand{\eeqn}{\end{eqnarray}}
\newcommand{\bi}{\bibitem}
\newcommand{\n}{\newline}
\newcommand{\nn}{\nonumber\\}
\maketitle

\begin{abstract}
We study the non-equilibrium relaxation of the spherical
spin-glass model with p-spin interactions in
the $N \rightarrow \infty$ limit.
We analytically solve the asymptotics of the
magnetization and the correlation and response functions for long
but finite times.
Even in the thermodynamic limit the system exhibits `weak' (as well as
`true') ergodicity
breaking and aging effects.
We determine a functional Parisi-like order parameter $P_d(q)$ which plays
a similar role for the dynamics to that played by
the usual function for the statics.

\vspace{2cm}
PACS numbers 02.50, 05.40, 64.60, 75.10
\end{abstract}

\newpage

\setcounter{page}{1}

Spin-glass dynamics has been a subject of continuous interest in the
past years.
Experimentally, spin-glass dynamics below the critical temperature
is characterized by aging effects and very slow relaxations \cite{Orbach}.
In long range mean-field models one knows that the phase space is broken
into ergodic components \cite{MPV}. Sompolinsky \cite{Sompo}
described a dynamics for these models allowing for barrier penetration
in very long times (diverging  as $N \rightarrow \infty $).

In realistic systems,  on the one hand mean-field is not exact and on the
other hand one cannot perform an experiment in infinite times, and one
actually sees at most `weak' ergodicity breaking.

Bouchaud has proposed a phenomenological scenario  with both `true'
and `weak' ergodicity breaking \cite{Bou}.
The question then arises as to if and how simple long-range microscopic
systems (for which mean-field is exact) can model these phenomena.
To the best of our
knowledge, an analytic description is lacking.

The main purpose of this paper is to show, in a very simple
mean-field model, the asymptotics of which we solve analytically,
that this is indeed so; in the thermodynamic limit `true' and `weak'
ergodicity
breaking coexist, and in a sense complement.
To this end we solve the dynamics of the p-spin spherical model ($p>2$)
first introduced in ref.\cite{CS}, {\it setting}
$N \rightarrow \infty$
{\it from the outset},  starting from a given
configuration,
for long (but not diverging with $N$) times.

It should be stressed that this is a different physical situation from
the Sompolinsky dynamics, which was analysed in ref.\cite{CHS}. We do not
have here any time-scale dependent on $N$ (or any other `regularization'
parameter): for the two-time ($t,t'$) functions the scale that naturally
arises is $t'/t$. Surprisingly, one can establish {\it formal}
contact with Sompolinsky's equations by defining a variable
$\tau = \log (t'/t)$ which plays the role of the `time' there
(this will be further explained in a separate
work in the context of the
Sherrington-Kirkpatrick (SK) model \cite{nos}).

Denoting the two-spin correlation and the linear response to a magnetic field
\[
C(t,t')
\equiv
\frac{1}{N} \sum_{i=1}^N \overline { \langle s_i(t) s_i(t') \rangle }
\;\;\;\;
G(t,t')
\equiv
 \frac{1}{N}
\sum_{i=1}^N
\frac{\partial \overline {\langle s_i(t) \rangle}}
     {\partial h_i(t')}
\; ,
\]
our main results are as follows:

{\it i.} For {\it any} waiting time
$t_w$ there exists a sufficiently large $t$ such that
$C(t+t_w,t_w)$ tends to zero.

{\it ii.} After a (large) $t_w$, the decay rate
of the correlation function has a short transient after which it is
inversely proportional to $t_w$ (an aging-like effect).

{\it iii.} For $t$ large, the magnetization falls to zero as $t^{-\nu}$.

{\it iv.} In addition to a strong short-term memory, the system possesses
a weak, long-term memory.

We expect other models, such as the Potts glass (for more than
three components) and the
p-spin Ising model (for not too low temperatures) to have a similar
dynamics to the one presented here.
The SK model instead has a rather
different behaviour \cite{nos}.

The spherical $p$-spin glass model is defined by the Hamiltonian
\beq
H =
-
\sum_{i_1 < \dots < i_p}^N
J_{i_1 \dots i_p}
s_{i_1} \dots s_{i_p}
+
\frac{1}{N^{r-1}}
\sum_{i_1 < \dots < i_r}^N
h_{i_1  \dots i_r}
s_{i_1} \dots s_{i_{r}}
\; .
\nonumber
\eeq
The spin variables verify the spherical constraint
$\sum_{i=1}^N s_i^2(t) = N$.
The interaction strenghts are
independent random variables with a Gaussian distribution with zero mean
and variance
$\overline{ (J_{i_1  \dots  i_p} )^2 }
=
p! /
(2 N^{p-1})
$.
The overline stands for the average over the couplings. Additional
source terms ($h_{i_1 \dots i_{r}}$ time-independent)
have been included; if $r=1$ the usual coupling to a magnetic
field $h_i$ is recovered.

The relaxational dynamics is given by the Langevin equation
\beq
\Gamma_0^{-1} \, \partial_t s_i(t)
=
- \beta \frac{\delta H}{\delta s_i(t)}
- z(t) s_i(t)
+ \xi_i(t)
\; .
\label{lang}
\eeq
$\Gamma_0$ determines the time scale and will be henceforth set to one.
The second term in the rhs enforces
the spherical constraint while
$\xi_i(t)$ is a Gaussian white noise
with zero mean and variance $2$.
The mean over the thermal noise is hereafter represented by
$\langle \;\cdot\;\rangle$.
As will be shown below,
the dynamical equations plus the spherical constraint impose
$z(t)=(1 - p \beta  {\cal E}(t))$ with ${\cal E}(t)$
the energy per spin.
We choose as initial configuration $s_i(0)=1$ $\forall i$,
though any other choice is equivalent.

The mean-field sample-averaged dynamics for $N \rightarrow \infty$
is entirely described by the evolution
 of the two-time  correlation
and the linear response functions. The dynamical equations for them
can be obtained
from eq.(\ref{lang})
through standard functional methods (see {\it eg.} ref.\cite{KT})
\beqn
\frac{\partial C(t,t')}{\partial t}
&=&
- \, (1- p \beta \,  {\cal E}(t)) \, C(t,t')
+
2 \, G(t',t)
\nn
& &
+
\,
\mu \int_0^{t'} dt'' \, C^{p-1}(t,t') \, G(t',t'')
\nn
& &
+
\,
\mu \, (p-1) \int_0^t dt'' \, G(t,t'') \, C^{p-2}(t,t'') \, C(t'',t')
\; ,
\label{corr}
\\
\frac{\partial G(t,t')}{\partial t}
&=&
- \,(1- p \beta \, {\cal E}(t)) \, G(t,t')
+ \delta(t-t')
\nn
& &
+
\,
\mu \, (p-1) \int_{t'}^t dt'' \, G(t,t'') \, C^{p-2}(t,t'') \, G(t'',t')
\; ,
\label{resp}
\eeqn
with $\mu \equiv p \beta^2 /2$.
These equations hold for all times $t$ and $t'$.
At equal times $C(t,t) = 1$,
$\lim_{t'\rightarrow t^{-}} G(t,t') = 1$
and
$\lim_{t'\rightarrow t^\pm} \partial_t C(t,t') = \pm 1$.
${\cal E}(t)$
can be identified as the energy per spin multiplying eq.(\ref{lang})
by $s_i(t')$, averaging over the noise and the couplings and taking
the limit $t'\rightarrow t$.
Furthermore, with the definition
\beq
I^r(t)
\equiv
\frac{r}{N}
\sum_{i_1 < \dots < i_r}
{
\frac{\partial
\overline{
<
s_{i_1}(t) \dots s_{i_r}(t)
>
}
}
{
\partial
h_{i_1 \dots i_r}
}
}\left|_{h=0} \right.
=
r
\int_0^t dt'' \, C^{r-1}(t,t'') \, G(t,t'')
\label{int}
\eeq
eq.(\ref{corr}) implies ${\cal E}(t) = (\beta/2) I^p(t)$.

\vspace{0.5cm}

In order to make the solution to  these equations intelligible, we first
briefly
describe the structure of the TAP free-energy landscape \cite{KPV}, though
we shall never use the TAP results in the dynamic treatment.
The TAP free-energy can be written in terms of $\hat s_i \equiv m_i/\sqrt q$
and
$q \equiv (1/N) \sum_i^N m_i^2 $ where
$m_i$ are the magnetizations
\[
f_{\rm TAP}= q^\frac{p}{2} \, {\cal E}_{0}(\hat s_1,\dots,\hat s_N)
- \frac{1}{2\beta} \ln (1-q)
- \frac{\beta}{4} \left[ (p-1) q^p-pq^{p-1}+1 \right]
\; .
\]
${\cal E}_{0}(\hat s_1,\dots,\hat s_N)$ denotes the {\em zero temperature}
energy of a configuration $\{ \hat s_i \}$.
The free-energy landscape in the `angular' variables
$\hat s_i$ is unaltered  by temperature apart from a stretching
proportional to $q^{p/2}$.

The `angular variable' saddle point equations are supplemented by the
condition of minimization with respect to $q$
\beq
q^{\frac{p}{2}-1} (1-q) = \frac{T}{p-1}
\left[ -{\cal E}_{0} + ( {\cal E}_{0}^2 - {\cal E}_{0c}^2)^{1/2} \right]
\label{dseis}
\eeq
where
${\cal E}_{0c} \equiv - (2(p-1)/p)^{1/2}$; the largest root for $q$
corresponds to the minimum.

The paramagnetic solution $q=0$ exists for all temperatures.
Each zero temperature saddle point $\{ \hat s_i \}$ with energy ${\cal
E}_0$ determines the temperature $T$
saddle point ($ m_i = \hat s_i \sqrt q$, $q$) where
$q$ is obtained from eq.(\ref{dseis}).
Thus, all TAP saddle points
are labelled by their associated zero temperature
energy ${\cal E}_{0}$.
Moreover, it is easy to see
that their ordering in free-energy does not change with temperature
({\it i.e.}
there is no `chaoticity' with respect to temperature in this model).

Above the threshold value ${\cal E}_{0c}$ for
${\cal E}_{0}$ (corresponding to a threshold for $f(\beta)$)
eq.(\ref{dseis}) has no solutions.
Since it will turn out that the dynamics is dominated by this threshold level,
it is useful to describe it in more detail.

Using standard methods \cite{BM} one
finds that
the typical spectrum of the free-energy
Hessian in a local minimum
corresponds to a `shifted'
semicircle law, with the lowest eigenvalue $\lambda_{min}$ given
(in terms of the parameters of the minimum) by
\[
\lambda_{min}=p \, q^{\frac{p}{2}-1} \, ({\cal E}_{0c} - {\cal E}_{0})
\; .
\]
Hence, for sub-threshold free-energies we have well-defined minima with no
`zero-modes' separated by $O(N)$ barriers. In particular, this was shown
within the replica approach for the lowest minima that dominate the
Gibbs-measure \cite{KPV}. Exponential decays would be expected within them;
however, those low-lying states are quite irrelevant
for the non-equilibrium dynamics of this model.
The gap $\lambda_{min}$ drops to zero at the threshold, and around that value
the barriers drop from $O(N)$ below to zero above.

The parameter $q$ and the TAP {\em energy} at the threshold are given by
\beqn
\frac{1}{p-1}
&=&
\mu \, q_{th}^{p-2} \, (1-q_{th})^2
\; ,
\label{tapq}
\\
{\cal E}_{th}
&=&
\frac{\beta}{2}
\left[
1 -
q^p_{th} (1-\frac{ (p-2) (1-q_{th})}{q_{th}})
\right]
\; .
\label{tapen}
\eeqn

\vspace{0.5cm}

Let us now turn to the solution of the dynamical equations.
Since we are interested in the non-equilibrium dynamics we solve
them with the only assumption of causality.
We take $t > t'$ for definiteness and
we focus on the {\it low} temperature phase.
The system (\ref{corr})-(\ref{resp}) can be solved numerically step by step
in a manner reminiscent of ref.\cite{opper}. The numerical solution
suggests the following scenario
for the asymptotic regime $t>>1$ which we later confirm analytically.
The time axis $t'$ is divided in three distinct zones with different
behaviours.

{\it i.}
If $t'$ is close to $t$ but $(t-t')/t \rightarrow 0$ asymptotically,
time homogeneity and the fluctuation-dissipation theorem (FDT) hold;
{\it i.e.}
$G_{\rm FDT}(t-t') =
-\Theta(t-t') \, \partial_t C_{\rm FDT}(t-t')$.
For large values of $(t-t')$
(but still small compared to $t$) $C_{\rm FDT}(t-t')$ tends to a
value $q$
and $G_{\rm FDT}(t-t')$ tends to zero.

{\it ii.}
If $t'$ is such that $(t-t')/t \sim O(1)$,
the relevant (adimensional) independent
variable turns out to be $\lambda \equiv t'/t$
($0<\lambda <1$).
In this sector the correlation and rescaled response functions
depend on $\lambda$ as $C(t,t') = q \, {\cal C}(\lambda)$ and
$t \, G(t,t') = {\cal G}(\lambda)$.
Since $q$ is the limiting value of $C(t,t')$ in the previous regime,
${\cal C}(1)=1$.

{\it iii.}
Finite times $t'$ correspond to $\lambda=0$ in rescaled variables. In
particular, for $t'=0$ we have the magnetization $m(t)=C(t,0)$.

\vspace{0.5cm}

We now proceed to solve the resulting equations within this asymptotic
scenario.
If $t'$ is such that the system is in the FDT regime
eq.(\ref{corr}) yields
\[
(\frac{\partial}{\partial t} + 1)
C_{\rm FDT}(t)
+
(\mu + p \beta \, {\cal E}_\infty)
\,
(1 - C_{\rm FDT}(t))
=
\mu \int_0^t dt''
\,
C^{p-1}_{\rm FDT}(t-t'') \,
\frac{dC_{\rm FDT}}{dt''}(t'')
\]
with the asymptotic energy ${\cal E}_\infty$ ({\it nb.} `$\infty$' is
understood as a limit taken {\it after} $N \rightarrow \infty$)
given by
\beq
{\cal E}_\infty
=
-
\frac{\beta}{2} \, \left[
(1-q^p)
+
p q^{p-1}
\int_0^1 d\lambda'' \, {\cal G}(\lambda'') {\cal C}^{p-1}(\lambda'')
\right]
\label{eninf}
\; .
\eeq
The correlation decays to a value $q$ determined by
\beq
1- p \beta \, {\cal E}_\infty + \mu \, (1-q^{p-1}) = -\frac{1}{1-q}
\label{qFDT}
\; .
\eeq
This equation appears in the dynamics \`a la Sompolinsky of
this model \cite{CHS}.
The solution for $q$ as well as the decay law
requires solving the coupled system
(\ref{eninf})-(\ref{qFDT}) which involves the previous history
through the $\lambda$-integration.

We now consider the regime $0<t'/t<1$.
The dynamic equations for this range of times
reduce to two coupled equations for ${\cal C}(\lambda)$ and
${\cal G}(\lambda)$ in which, consistently, all times enter {\em only}
through $\lambda$:
\beqn
0
&=&
{\cal G}(\lambda)
\left [
-(1-q)^{-1}
+
\mu \, (1-q) (p-1) \, q^{p-2} \, {\cal C}^{p-2}(\lambda)
\right ]
\nn
& &
+
\mu \, (p-1) \, q^{p-2}
\int_\lambda^1
\frac{d\lambda''}{\lambda''} \, {\cal G}(\lambda'')
\, {\cal C}^{p-2}(\lambda'')
\, {\cal G}\left(\frac{\lambda}{\lambda''}\right)
\; ,
\label{gcursiva}
\\
0
&=&
{\cal C}(\lambda)
\left [
-(1-q)^{-1} + \mu \, (1-q) q^{p-2} {\cal C}^{p-2}(\lambda)
\right ]
\nn
& &
+
\mu \, q^{p-2}
\int_0^\lambda
\frac{d\lambda''}{\lambda} \, {\cal C}^{p-1}(\lambda'') \,
{\cal G} \left(\frac{\lambda''}{\lambda}\right)
\nn
& &
+
\mu \, (p-1) \, q^{p-2}
\int_0^1
d\lambda'' \, {\cal G}(\lambda'') \, {\cal C}^{p-2}(\lambda'') \,
{\cal C} \left( (\frac{\lambda}{\lambda''})^{{\rm sgn}(\lambda''-\lambda)}
\right)
\label{ccursiva}
\; .
\eeqn
Eq.(\ref{gcursiva}) in $\lambda=1$ admits the solution ${\cal G}(1)=0$
which implies ${\cal G}(\lambda)\equiv 0$ and this is the {\it high}
temperature
asymptotics.
In the {\it low} temperature phase a non-trivial
${\cal G}(\lambda)$ is possible provided the first square bracket in
(\ref{gcursiva}) evaluated in $\lambda=1$ is zero; this fixes the value $q$.
{}From eq.(\ref{gcursiva}) it also follows ${\cal G}(1)= x \,q \,{\cal C}'(1)$
(prime denotes derivative with respect to $\lambda$)
$x\equiv (p-2)(1-q)/q$.
It is now easy to
see that the system (\ref{gcursiva})-(\ref{ccursiva}) with
${\cal G}(\lambda) = x \, q \, {\cal C}'(\lambda)$
simplifies to a single equation. With this ansatz the system of equations
has the unique family of (exact) solutions
\beqn
\begin{array}{rclcrcl}
{\cal  C}(\lambda) &=& \lambda^\nu &
\Longleftrightarrow &  C(t,t') &=& q \left( \frac{t'}{t} \right)^\nu
\; .
\nonumber
\end{array}
\eeqn
In order to determine $\nu$ ($0<\nu<1$) a careful
matching between this solution and the ones associated with
other sectors has to be made.

All the integrals (\ref{int}) in the large $t$ limit become
\beq
I^r_\infty = 1 - q^r (1-x)
\label{intinf}
\; .
\eeq
In particular, ${\cal E}_\infty = (\beta/2) I^p_\infty$.
We have now everything that is required to solve the FDT relaxation,
eqs.(\ref{eninf})-(\ref{qFDT}), which {\it for this  value of}
${\cal E}_\infty$
imply a power law decay for this regime \cite{CHS}.
Interestingly enough, the expressions just derived for
the energy ${\cal E}_\infty$ and $q$ coincide with
eqs.(\ref{tapq})-(\ref{tapen}).
Therefore, we have learned that
the long time dynamics takes place in the  threshold of the TAP free-energy.

Finally, we consider the finite $t'$ regime. We already know that for
large $t$ correlations relax to zero; we now study the asymptotics.
Inserting the behaviour $C(t,t')\sim t^{-\alpha} c(t')$ in
eqs.(\ref{corr})-(\ref{resp}) and using the previous results, we find
$\alpha=\nu$, {\it i.e.} the exponent for $t$ is the same as in the
previous regime.

The numerical solutions show that
the asymptotic regime is well established  already
for (adimensional) times $t\sim 100$.

\vspace{0.5cm}

We have hence the following picture. For {\it low}
temperature and increasing times the system
explores deeper and deeper traps, the permanence time in a trap being
small compared with $t$. This allows for an equilibration between a few
traps at every stage. From the asymptotic solutions and definition
(\ref{int}) we have
\beq
1-
\frac{r}{N^r}
\sum_{i_1 < \dots < i_r}
{
\frac{\partial
\overline{
<
s_{i_1}(t) \dots s_{i_r}(t)
>
}
}
{
\partial
h_{i_1 \dots i_r}
}
}\left|_{h=0} \right.
=
\int_0^1 dq' \, P_d (q') \, {q'}^r
\label{aa}
\eeq
with $P_d (q')$ tending to ({\it cf.} eqs.(\ref{int}) and (\ref{intinf}))
\beq
P_d (q') = x \, \delta(q') + (1-x) \, \delta(q'-q)
\label{bb}
\; .
\eeq
Assuming a Boltzmann distribution
restricted to the visited region of phase space at any stage, and that
the long-time traps tend to verify clustering,
starting from
eqs.(\ref{aa})-(\ref{bb})
we can use similar arguments to those leading to the interpretation
of the static $P(q)$ \cite{MPV} in order to understand the dynamical
$P_d(q)$ (note that in this model $P_d(q) \neq P(q)$).
Much of the interpretation of the static $P(q)$ carries on
to the measure associated to the pseudo-equilibria. There is however
an important difference: the identity of the dominating {\it pseudo}-states
changes with time (otherwise these would be {\it bona fide} states which
we have previously seen they are not). The dynamic phase transition takes place
when $x$ reaches one in a manner that resembles
the static transition \cite{KPV};
at that point the threshold energy coincides with
the paramagnetic energy.

It can come as a surprise that analytic results can be obtained at all
in such non-equilibrium situations: the underlying reason is the weakness
of the memory of the system. It would be interesting to understand whether
this also holds for more realistic systems.

\vspace{1cm}

We wish to acknowledge useful discussions with A. Crisanti,
and H. Rieger. We are indebted with
S. Franz, E. Marinari, G. Parisi and M.A. Virasoro
for critical reading of the manuscript and suggestions.

\end{document}